\documentclass[prb, %twocolumn,
superscriptaddress,showpacs,amsmath,amssymb]{revtex4}
\usepackage{amsfonts}
\usepackage{bm}
\usepackage{verbatim}
\usepackage{graphicx}

\graphicspath{{pics/}}

\begin{document}

\title{Chiral vortical effect generated by chiral anomaly in vortex-skyrmions}

\author{G.E.~Volovik}
\affiliation{Low Temperature Laboratory, Aalto University,  P.O. Box 15100, FI-00076 Aalto, Finland}
\affiliation{Landau Institute for Theoretical Physics RAS, Kosygina 2, 119334 Moscow, Russia}

%\pacs{67.30.H-, 14.80.Bn, 67.30.hj, 67.85.Fg}

\date{\today}

\begin{abstract}
  We discuss the type of the general macroscopic parity-violating effects, when there is the current along the vortex, which is concentrated in the vortex core.  We consider vortices in chiral superfluids with Weyl points. In the vortex core the positions of the Weyl points form the skyrmion structure. We show that the mass current concentrated in such a core is provided by the spectral flow through the Weyl points according to the Adler-Bell-Jackiw equation for chiral anomaly. 
\end{abstract}

\maketitle

 \section{Introduction}

The problems with Weyl materials -- Weyl semimetals and Weyl superconductors (see latest reviews in Ref. \cite{Hasan2016} and Ref. \cite{Sato2016} correspondingly) -- is that at first glance they should possess any type of a bulk response that
exists in conventional non-Weyl materials with the same symmetry \cite{Pesin2016,Pesin2015}. 
That is why the task is to resolve the contribution of the anomalies, which accompany the physics of Weyl fermions, from the conventional contributions, which follow from the symmetry consideration.  Here we consider the situation when the effect is fully determined by chiral anomaly. This is the chiral vortical effect (CVE) produced by the vortex-skyrmions in Weyl superfluids/superconductors.
 
   \section{CVE by skyrmion in chiral superfluid with two Weyl nodes}

  For simplicity we consider the Galilean invariant system, where the mass current coincides with the momentum density, and in case of charged particles the electric current coincides with the mass current
 with the factor $e/m$.

%%%%%%%%%%%%%%%%%%%%%%%%%%%%%%%%%%%%%%%%
%%%%%%%%%%%%%%%%%%%%%%%%%%%%%%%%%%%%%%%%
\begin{figure}[top]
\centerline{\includegraphics[width=0.8\linewidth]{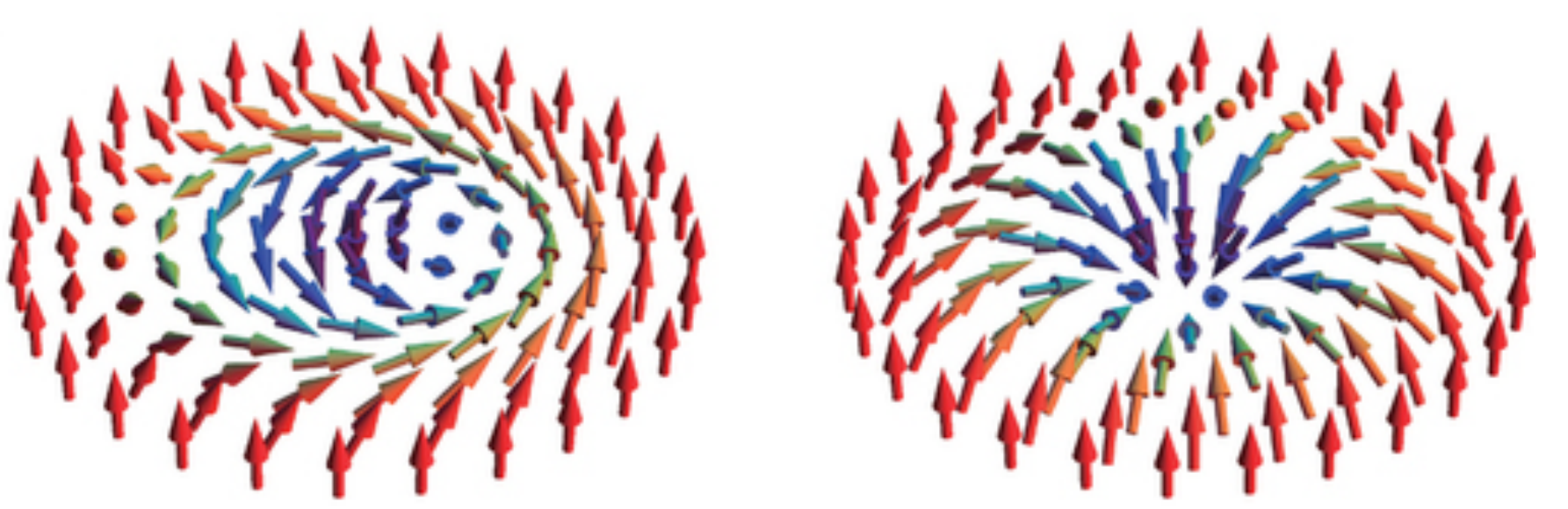}}
\label{Skyrmions} 
  \caption{ {Fig. 1 \it Right}: Neel-type vortex skyrmion, which
is symmetric under the combined symmetry $PTO_{x,\pi}$, which forbids the current along the vortex axis, since $J_z=PTO_{x,\pi}J_z= - J_z$.  {\it Left}:  Bloch-type skyrmion, which
is symmetric under the combined symmetry $TO_{x,\pi}$. The broken $P$-symmetry allows the current along the vortex axis -- the chiral vortical effect. The main difference from the Neel and Bloch skyrmions discussed in the nonsuperconducting magnetic materials in Ref. \cite{Kezsmarki2015} is that in the chiral superfluids there is the circulation of superfluid velocity around the skyrmion. In the $p$-wave superfluids the skyrmion represents the vortex with two quanta of circulation \cite{SalomaaVolovik1987}.
 }
\end{figure}
%%%%%%%%%%%%%%%%%%%%%%%%%%%%%%%%%%%%%%%%
%%%%%%%%%%%%%%%%%%%%%%%%%%%%%%%%%%%%%%%%

We discuss a particular type of the parity violating effects \cite{Vilenkin1979,Vilenkin1980,Kharzeev2015}, when there appears the current along the vortex axis, which is concentrated in the vortex core. As an example of vortices, which experience such a chiral vortical effect (CVE), we consider vortex-skyrmions in Weyl superfluids.
The vortex-skyrmion is the continuous (non-singular) texture in Fig. 1,
which has the skyrmion structure of the orbital magnetization, and the superflow with two quanta of circulation around the skyrmion, see review \cite{SalomaaVolovik1987}. 
For such inhomogeneous configuration the traditional response theory, which results depend on the order of limits $q\rightarrow 0$ and $\omega\rightarrow 0$ \cite{Pesin2016}, is not applicable. Instead we can use the trick, which
has already been used in chiral Weyl superfluids for the calculation of the angular momentum of the texture  \cite{Volovik1995} (see also Refs. \cite{Tada2015,Volovik2014}), and for the calculation of different manifestations of the chiral magnetic effect (CME) \cite{Volovik2017a}.
We calculate the current generated by deformation of the order parameter within the skyrmion, when 
the texture is deformed from the state obeying the space inversion $P$ to the state with the violated space inversion symmetry.
In the initial state there is no current along the axis of the skyrmion, since it is forbidden by the $P$-symmetry. The task is to find the current in the final sate using the spectral flow in the process of deformation. We shall show that in this process the CVE current emerges and it fully originates from the chiral anomaly.

If the texture has no dependence on the coordinate $z$ along the vortex-skyrmion, then  in the process of deformation of the texture there is no force applied in $z$ direction. 
In this situation the change of the total linear momentum or equivalently of the mass current $J_z$ may come only from the spectral flow from the occupied negative energy states of the inhomogeneous vacuum of skyrmion to the positive energy world, where the momentum of the created particles is accumulated. Such spectral flow may take place either through the nodes in the bulk spectrum \cite{Volovik1995},  or through the boundaries \cite{Tada2015,Volovik2014}.
Since the skyrmion is the localized object, which is not connected to the side walls of the cylindrical container, the boundary effects can be ignored. Then, if the bulk state is fully gapped,
there will be no current along the skyrmion axis, even if the space inversion (or the combined space inversion) is broken. But in the materials with Weyl point nodes in bulk, the spectral flow through the Weyl nodes results in the total current, which is fully regulated by the Adler-Bell-Jackiw equation \cite{Adler1969,Adler2005,BellJackiw1969} for the chiral anomaly experienced by the Weyl fermions.

We consider the system, which has two Weyl points with opposite chiralities, but the extension to the superfluids with several Weyl points is straightforward \cite{Volovik2017b}.
Such Weyl nodes cannot occur in the Galilean invariant normal state, but may occur in the pair-correlated gases and liquids, such as the A-phase of superfluid $^3$He.  The latter is the spin-triplet chiral $p$-wave superfluid, where the Weyl points are at ${\bf K}_\pm= \pm k_F \hat{\bf l}$, and $\hat{\bf l}$ is the unit vector in the direction of the angular momentum of Cooper pairs. 

We start with the axi-symmetric texture of the vortex-skyrmion. In the texture, the positions of Weyl nodes in momentum space are the following functions of the coordinates: 
 \begin{equation}
{\bf K}_\pm ({\bf r})=\pm k_F\left(\hat{\bf z} \cos\eta(r) +\sin\eta(r) (\hat{\bf r}\cos\alpha + 
\hat{\mbox{\boldmath$\phi$}} \sin\alpha)\right).
\label{NodePosition}
\end{equation}
Here $(z,r,\phi)$ are cylindrical coordinates;  $\eta(r)$ changes from $\pi$ to $0$ ($\eta(0)=\pi$ and $\eta(\infty)=0$); the parameter $\alpha$ is constant. For $\alpha=0$, the texture is shown in Fig. 1  ({\it right}). It is the  Neel-type skyrmion, which
is symmetric under the combined symmetry $PTO_{x,\pi}$, where  $P$ is space inversion, $T$ is time inversion, and $O_{x,\pi}$ is $\pi$ rotation about horizontal axis.  This symmetry forbids the current along the vortex axis, since $J_z=PTO_{x,\pi}J_z= - J_z$.  For $\alpha\neq 0$ this  symmetry is violated and the current along the vortex axis may exist. For $\alpha =\pi/2$ it is the Bloch-type skyrmion  in Fig. 1, 
 which is symmetric under the combined symmetry $TO_{x,\pi}$.  Neel and Bloch skyrmions in nonsuperconducting magnetic materials see e.g. in Ref. \cite{Kezsmarki2015}.

Here we show that for $\alpha\neq 0$ the current along the vortex comes from the chiral anomaly experienced by the Weyl fermions, and it vanishes when the Weyl nodes annihilate each other.
Let us consider deformation, at which $\alpha(t)$ changes from zero to the finite value. The positions of the Weyl points play the role of the effective vector potentials ${\bf A}_\pm({\bf r},t) ={\bf K}_\pm({\bf r},t) $ acting on fermions in the vicinity of these two Weyl points (in superconductors such synthetic gauge fields are induced by strain \cite{Pikulin2016}). This results in the effective electric and magnetic  fields:
\begin{equation}
{\bf E}_\pm =-\partial_t {\bf K}_\pm \,\, , \,\, {\bf B}_\pm =\nabla \times {\bf K}_\pm
\,,
\label{EMfields}
\end{equation}
with
\begin{equation}
{\bf E}_+ \cdot {\bf B}_+= {\bf E}_-\cdot {\bf B}_- =-k_F^2 \frac{d \cos \eta}{dr} \sin \eta \frac{d \sin \alpha}{dt}
\,.
\label{BE}
\end{equation}
The production of the chiral charge at two Weyl points due to chiral anomaly compensate each other:
\begin{equation}
\dot n_5 =\frac{1}{4\pi^2} \left({\bf E}_+ \cdot {\bf B}_+ - {\bf E}_-\cdot {\bf B}_- \right)=0
\,,
\label{dotn}
\end{equation}
But the created chiral charge carries with it the linear momentum ${\bf K}_\pm$, which is not cancelled. As a result there is the momentum production per unit time per unit volume:
\begin{eqnarray}
\dot j_z=\frac{1}{4\pi^2}\left( K_{z+}({\bf E}_+ \cdot {\bf B}_+ ) -K_{z-}( {\bf E}_-\cdot {\bf B}_-) \right)= 
\nonumber
\\
-\frac{k_F^3}{2\pi^2}  \cos\eta\frac{d \cos \eta}{dr} \sin \eta \frac{d \sin \alpha}{dt}
\,.
\label{dotP}
\end{eqnarray}
The linear momentum accumulated by the vortex in the process of deformation, and thus the mass current along the skyrmion,  is obtained by integration over time and space:
\begin{eqnarray}
J_z({\rm vortex})=\int d^2 r \int dt \,\dot j_z=
\nonumber
\\
-\frac{k_F^3}{2\pi^2} \sin \alpha\int d^2r  \cos\eta\frac{d \cos \eta}{dr} \sin \eta 
= 
\nonumber
\\
-\frac{k_F^3}{6\pi} \sin \alpha\int_0^\infty dr \, \sin^3 \eta 
  \,.
\label{Jz}
\end{eqnarray}

 \section{CVE in $^3$He-A}

%%%%%%%%%%%%%%%%%%%%%%%%%%%%%%%%%%%%%%%%
%%%%%%%%%%%%%%%%%%%%%%%%%%%%%%%%%%%%%%%%
\begin{figure}[top]
\centerline{\includegraphics[width=0.6\linewidth]{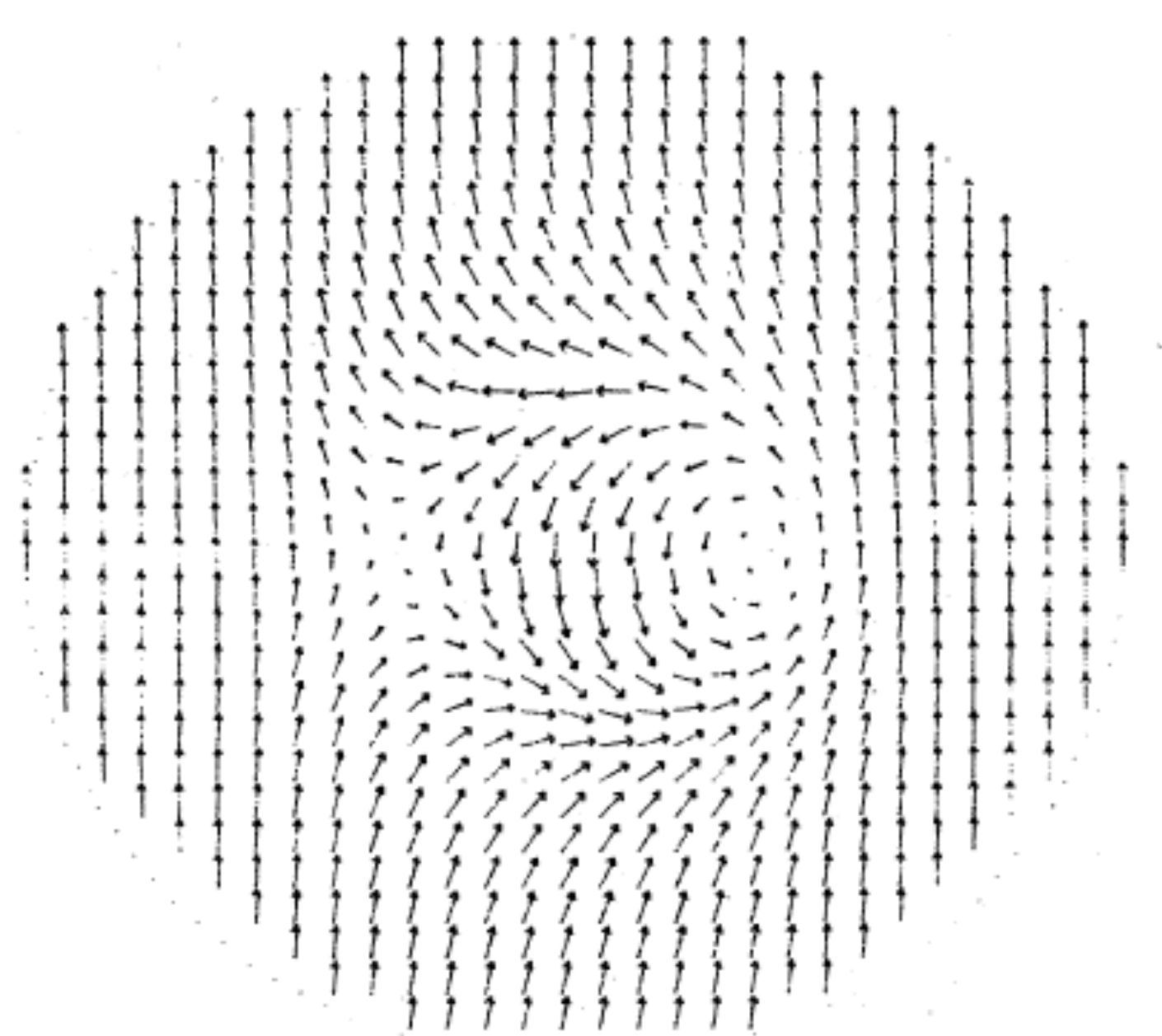}}
\label{wvortex} 
  \caption{The cross section of the $w$-vortex skyrmion in superfluid $^3$He-A (from Ref. \cite{SalomaaVolovik1987}). The texture is symmetric under the combined symmetry $TO_{x,\pi}$, while the parity $P$ is broken. The broken $P$-symmetry in the skyrmion allows the current along the vortex axis in Eq.(\ref{JzW}) -- the chiral vortical effect.  As distinct from the Bloch-type skyrmion in Fig. 1 ({\it left}), this vortex is not axisymmetric and consists of two merons. Nevertheless its chiral vortical effect is also fully determined by chiral anomaly.
 }
\end{figure}
%%%%%%%%%%%%%%%%%%%%%%%%%%%%%%%%%%%%%%%%
%%%%%%%%%%%%%%%%%%%%%%%%%%%%%%%%%%%%%%%%

Let us compare the value of the mass current in Eq.(\ref{Jz}) with the current in the core of the vortex skyrmion in superfluid $^3$He-A in  Eq.(5.35) of review \cite{SalomaaVolovik1987}. The vortex-skyrmion with non-zero current is the so-called $w$-vortex with $TO_{x,\pi}$ symmetry shown in Fig. 2.
Its structure is given by Eq.(5.29) in Ref. \cite{SalomaaVolovik1987}:
  \begin{equation}
{\bf K}_\pm ({\bf r})=\pm k_F\left(\hat{\bf y} \cos\eta(r) +\sin\eta(r) (\hat{\bf x}\sin\phi -  
\hat{\bf z} \cos\phi)\right).
\label{NodePositionW}
\end{equation}
This $w$-vortex is analogous to the skyrmion with $\alpha=\pi/2$ in Eq.(\ref{NodePosition}), but as distinct from the vortex in Eq.(\ref{NodePosition}),
the texture of the $\hat{\bf l}$-vector is not axisymmetric: this vector changes from $\hat{\bf l}(r=0)= -\hat{\bf y}$ to $\hat{\bf l}(r=\infty)= \hat{\bf y}$.
The reason for such asymmetry is that is that the NMR experiments are made in applied magnetic field, and  far from the vortex core the  $\hat{\bf l}$-vector is oriented perpendicular to the field direction due to spin-orbit interaction.

The current along the $w$-vortex axis \cite{SalomaaVolovik1987} is expressed in terms of the hydrodynamic parameters calculated by Cross \cite{Cross1975} from the microscopic BCS theory:
\begin{eqnarray}
J_z({\rm vortex})=  
\nonumber
\\
-\frac{\pi}{4m} \int_0^\infty dr \left( C_0\sin^3 \eta
+(\rho_s^\perp-\rho_s^\parallel)(1+\cos\eta) \sin\eta \cos\eta\right).
\label{JzW}
\end{eqnarray}
Here $m$ is the mass of the $^3$He atom, $\rho_s^\parallel$ and $\rho_s^\perp$ are superfluid densities for the superfluid motion along and perpendicular to the $\hat{\bf l}$-vector. 

At $T=0$ the superfluid density tensor becomes isotropic, $\rho_s^\parallel(T=0)=\rho_s^\perp(T=0)=\rho$,  and the hydrodynamic parameter $C_0$ approaches the value
$C_0(T=0)=mk_F^3/3\pi^2$. As a result the Eq.(\ref{JzW}) is reduced to Eq.(\ref{Jz}) 
for the current in the axisymmetric skyrmion with  $\alpha=\pi/2$.
Note that in the traditional approach \cite{Cross1975} the hydrodynamic parameters have been calculated using fermionic spectrum very far from the nodes. Nevertheless, the CVE current is fully determined by the spectral flow through the nodes. This demonstrates the universality of the contribution of the Weyl fermions to the CVE. 

It is interesting that the CVE current in Eq.(\ref{Jz}) is determined solely by the positions of the nodes and do not depend on any other parameter of the system, including the mass $m$ of the atom. For the considered Ansatz (\ref{NodePosition}) one has:
\begin{equation}
J_z({\rm vortex})=\frac{1}{6\pi \hbar ^2}\int_0^\infty  dr\, K^2_\perp(r)K_\phi(r)
  \,,
\label{CVE3HeA}
\end{equation}
where
 \begin{equation}
 K^2_\perp(r)=|{\bf K}|^2 \sin^2(r)
\,\,, \,\, K_\phi(r) =|{\bf K}| \sin\alpha\sin\eta(r)\,.
\label{NodeContribution}
\end{equation} 
The current disappears when the Weyl nodes merge at $|{\bf K}|\rightarrow 0$ and annihilate each other. This means that the CVE current is fully determined by the chiral anomaly experienced by the Weyl fermions.

Using the spectral asymmetry arguments  \cite{Volovik1995,Tada2015,Volovik2014} one can obtain the CVE current not only in vortex-skyrmions, but also in singular vortices with broken parity. The spectrum of fermions in the axisymmetric vortex $E(p_z,Q)$ depends on two quantum numbers, the linear momentum $p_z$ and angular momentum $Q$. Typically the minigap is much smaller than the gap in bulk, and the quantization of $Q$ can be ignored. In this case the current concentrated in the vortex core (the total linear momentum of the vortex) is
\begin{equation}
J_z({\rm vortex})=\sum_{p_z,a} p_z Q_a(p_z) \,.
\label{CVEvortex}
\end{equation}
Here $Q_a(p_z)$ are the values of $Q$, at which the spectrum crosses zero energy level as a function of $Q$ at fixed $p_z$, i.e. these are the solutions of equation $E(p_z,Q)=0$.
 
Note that in the ground state of the system or in the equilibrium state in general, the total current is absent due to the Bloch theorem, see e.g. Ref. \cite{Yamamoto2015}. In our case the current $J_z({\rm vortex})$ within the skyrmion  in Eqs.  (\ref{Jz}) and (\ref{JzW}) or within the vortex core in Eq.(\ref{CVEvortex}) in general, is compensated by the bulk current $J_z({\rm bulk})=\int d^2r\,\rho v_{sz} $\cite{SalomaaVolovik1987}. This can be seen from the following consideration. The equilibrium or the ground state of the system corresponds to the energy  minimum of the superfluid with  skyrmion.  The relevant part of the free energy contains two terms --  the energy of the superflow along $z$ in bulk liquid and the interaction of the bulk superflow with the current inside the vortex:
\begin{equation}
 F= \frac{1}{2}\int d^2r \rho v^2_{sz} +
v_{sz}  J_z({\rm vortex}) \,.
\label{CVE3HeA}
\end{equation}
The ground state is determined by minimization of $F$ with respect to the superfluid velocity $v_{sz}$ of the bulk flow, $dF/dv_{sz}=0$ \cite{SalomaaVolovik1987}. On the other hand the variation of the energy over $v_{sz}$ is nothing but the total current along $z$, i.e. $dF/dv_{sz}= J_z({\rm total})$. That is why the total current is zero in equilibrium, $J_z({\rm total})=J_z({\rm bulk})+ J_z({\rm vortex})=0$. For a single vortex in a large volume of the bulk liquid, the density of the compensating bulk current is much smaller than the current density in the vortex core, $j_z({\rm bulk})=  (R_{\rm core}^2/R^2) \, j_z({\rm vortex}) \ll j_z({\rm vortex})$. Here $R_{\rm core}$ is the core size of vortex or skyrmion, and $R$ is the radius of the container.

The  Bloch theorem is also the reason, why the equilibrium CVE is not possible in the normal (non-superfluid) state of the liquid: there is no non-dissipative supercurrent in bulk which could compensate the CVE current, and thus  in the ground state one has $J_z({\rm vortex})=0$. The same refers to the chiral magnetic effect (CME), where it was shown that there is no current in response to a strictly static magnetic field \cite{Chen2013,Zubkov2016}. In superconductors the non-uniform current along the magnetic field is possible if parity is properly violated, but again the total current is zero in equilibrium \cite{Levitov1985,Nazarov1986}.
In superconductors with broken parity  it is difficult to resolve between the conventional CME and the CME originating from the Weyl nodes, but in some cases the Weyl contribution can be dominating
\cite{Beenakker2016}.

 \section{Conclusion}

Superfluids, which contain the Weyl points, experience the type of chiral vortical effect, when there is the current along the vortex, which is concentrated in the vortex core. We considered the vortices, which cores have the skyrmion structure. The current concentrated in such a core is provided by the spectral flow through the Weyl points according to the Adler-Bell-Jackiw equation for chiral anomaly. In a full equilibrium the current along the vortex is compensated by the superfluid counter-current in the bulk superfluid outside the vortex core.

I thank M. Zubkov for numerous discussions.
This project has received funding from the European Research Council (ERC)
under the European Union's Horizon 2020 research and innovation programme
(Grant Agreement \# 694248).

\end{document}